\documentstyle[12pt]{article}
\newcommand{\be}{\begin{equation}}
\newcommand{\ee}{\end{equation}}
 
\newcommand{\bea}{\begin{eqnarray}}
\newcommand{\eea}{\end{eqnarray}}
\newcommand{\ba}{\begin{array}}
\newcommand{\ea}{\end{array}}
\newcommand{\beqa}{\begin{eqnarray}}
\newcommand{\eeqa}{\end{eqnarray}}

\newcommand{\lsim}
{{\;\raise0.3ex\hbox{$<$\kern-0.75em\raise-1.1ex\hbox{$\sim$}}\;}}
\newcommand{\gsim}
{{\;\raise0.3ex\hbox{$>$\kern-0.75em\raise-1.1ex\hbox{$\sim$}}\;}}

\newcommand{\PR}[1]{Phys. Rev.\ { #1}\ }

\newcommand{\Tr}{{\rm Tr}}

\newcommand{\third}{\displaystyle\frac{1}{3}}

\newcommand{\ssu}{$SU(2)_L\times SU(2)_R\times U(1)_{B-L}\,$}

\newcommand{\sulu}{$SU(2)_L\times U(1)_Y$}

\newcommand{\matr}{\left( \begin{array}}
\newcommand{\ematr}{\end{array} \right)}

\newcommand{\dis}{\displaystyle}

\begin{document}

\begin{flushright}
\parbox[t]{4.8truecm}{
\begin{center}
{
HU-SEFT R 1996-16, FTUV 96/34, IFIC 96/40\\}
{(hep-ph/9606311)}
\end{center}}
\end{flushright}
\vskip 1.5cm

\centerline{\bf DOUBLY CHARGED HIGGS AT LHC}
\vspace{0.8cm}
\centerline{\rm K. HUITU}
\baselineskip=13pt
\centerline{\it Research Institute for High Energy Physics}
\centerline{\it  University of Helsinki, Finland }
\vspace{0.3cm}
\centerline{\rm J. MAALAMPI}
\baselineskip=13pt
\centerline{\it Department of  Physics, Theory Division }
\centerline{\it University of Helsinki, Finland }
\vspace{0.3cm}
\centerline{\rm A. PIETIL\"A}
\baselineskip=13pt
\centerline{\it Department of Applied Physics }
\centerline{\it  University of Turku, Finland }
\vspace{0.3cm}
\centerline{\rm and}
\vspace{0.3cm}
\centerline{\rm M. RAIDAL}
\baselineskip=13pt
\centerline{\it Department of Theoretical Physics }
\centerline{\it  University of Valencia, Spain }
\vspace{0.9cm}
\abstract{{ 
We have investigated  production of doubly charged Higgs particles
$\Delta_{L,R}^{++}$ via  $WW$ fusion process
in proton-proton collisions at LHC energies in the framework of 
 the left-right symmetric model.  
The production cross section of the right-triplet Higgs 
$\Delta_R^{++}$ is for representative
values of model parameters at femtobarn level. The discovery reach
depends on the mass of the right-handed gauge boson $W_R$.
At best $\Delta_R^{++}$ mass up to
2.4 TeV are achievable within one year run. 
For  $\Delta^{++}_L$ the corresponding limit is 1.75 TeV which depends on 
the value of the left-triplet vev $v_L.$
Comparison with Drell-Yan pair production processes shows that
studies of the $WW$ fusion processes extend the discovery 
reach of LHC roughly by a factor of two.  
The main experimental signal of a produced
$\Delta^{++}_{L,R}$ would be a hard same-sign lepton pair. There will be no
substantial background due to the Standard Model (SM) interactions,
since in the SM
 a same-sign lepton pair will always be associated with
missing energy, i.e. neutrinos,  due to lepton number conservation.
}}

\bigskip
 
\vfil
\rm\baselineskip=20pt

\section{Introduction}

Doubly charged scalar particles arise in many scenarios 
\cite{cuy} extending the 
weak interactions beyond the Standard Model (SM). In the 
 left-right symmetric (LR)  electroweak theory \cite{pati} such a
particle   is a member of a triplet Higgs representation which plays a
crucial part in the model. The gauge symmetry  \ssu\ of the LR model  is
broken to the SM symmetry
\sulu\  due to a triplet Higgs $\Delta_R$, whose neutral component
acquires a non-vanishing expectation value in the vacuum. The
$\Delta_R$, called the right-triplet, transforms according to
$\Delta_R=(1,3,2)$, and it consists of the complex fields
$\Delta^0_R$, $\Delta^+_R$ and $\Delta^{++}_R$. If  the Lagrangian is
assumed to be invariant under a discrete $L\leftrightarrow R$  symmetry,
it must contain, in addition to $\Delta_R$, also a left-triplet
$\Delta_L=(\Delta^0_L,\Delta^+_L,\Delta^{++}_L)=(3,1,2)$. Hence the LR
model predicts two kinds of doubly charged particles with different
interactions. In contrast with $\Delta_R$, the existence of $\Delta_L$
is not  essential from the point of view of the spontaneous symmetry
breaking of the gauge symmetry. The vacuum expectation value $v_L$ of
its neutral member is actually quite  tightly bound by the $\rho$
parameter, i.e. by the measured  mass ratio of the ordinary weak bosons.

 The triplet Higgses have the following Yukawa couplings to the 
 leptons:
\be {\cal L}_Y= h_{R,ij}\Psi_{iR}^TC\sigma_2\Delta_R\Psi_{jR} + 
h_{L,ij}\Psi_{iL}^TC\sigma_2\Delta_L\Psi_{jL}\;\;\;+\;{\rm h.c.},
\label{yukawa}
\ee where $\Psi_{iR,L}=(\nu_{iR,L},l_{iR,L})$ and $i,j$ are flavour
indices. From the point of view of phenomenology a very important fact
is that the U(1)$_{B-L}$ symmetry prevents  quarks from coupling to
$\Delta_R$ and $\Delta_L$. In the processes that involve hadrons the
triplet Higgses appear thus only through higher order corrections.

The Yukawa Lagrangian (\ref{yukawa}) leads to large Majorana mass terms
of the form $h_{R,ij}\langle\Delta^0_R\rangle\nu_{iR}\nu_{jR}$ for the
right-handed neutrinos. These give rise to the see-saw mechanism
\cite{seesaw}, which provides the simplest explanation to the lightness
of ordinary neutrinos, if neutrinos do have a mass. The anomalies measured in
the solar
\cite{sun} and atmospheric
\cite{atmos} neutrino fluxes seem  to require that neutrinos indeed 
have a
 mass, manifested in these phenomena through flavour oscillations.
Furthermore, the  observations of  COBE satellite
\cite{cobe} may indicate the existence  of   a hot neutrino component
 in the dark matter of the Universe. In the framework of SM these
observations are difficult to explain, since  there neutrinos are predicted
to be massless.

Apart from the question of neutrino mass, the LR model is more
satisfactory than the SM also in that it gives a better understanding of
parity violation and it maintains the lepton-quark symmetry  in weak
interactions.

Nevertheless, so far there has been no direct evidence of  left-right
symmetry  in weak interactions. This sets a lower bound to the energy
scale of the breaking of that symmetry. According to the direct searches
of the CDF and D0 experiments at Tevatron, the intermediate bosons of
the right-handed interactions have the mass limits 
$M_{W_R}\gsim 652$ GeV \cite{CDF} and $M_{W_R}\gsim 650$ GeV (720 GeV if
the right-handed neutrino is assumed to be much lighter than $W_R$)
\cite{D0}, respectively. The mass limit for the new neutral intermediate
boson is
$M_{Z_2}\gsim 445$ GeV \cite{PDG}. Although  there are some assumptions
behind these bounds concerning e.g. the  strength $g_R$ of the
right-handed gauge interactions in comparison with the strength $g_L$ of
the left-handed interactions
 and the form of the CKM matrix of the right-handed interactions, which
may, when relaxed,  degrade the bounds considerably
\cite{rizzoap95}, it is reasonable to assume that below the scale of 0.5 TeV
the left-right symmetry is broken.

Hence, if the right-handed electroweak interactions exist, their
discovery would require accelerators whose capacity exceeds that of the
present ones. The phenomenological signatures of the LR model in  0.5
-- 2 TeV linear colliders have been recently under intensive study 
\cite{NLC}. Particularly the signatures of the doubly charged scalar at
$e^+e^-$ linear collider were discussed in ref. \cite{gunionDelta}. The
production of a single doubly charged Higgs in $ep$ collisions at Hera was
studied in ref. \cite{acco}.

In previous works only the pair production of the left-triplet 
$\Delta^{++}_L$ in $pp$ collisions at SSC has been investigated 
\cite{grifols}. 
In the present paper we shall investigate  the prospects of testing  the
production of single doubly charged Higgs scalars in high-energy
$pp$ collisions at LHC. Our main concern will be the right-triplet
$\Delta^{++}_R$ but we will also study the production of the
left-triplet $\Delta^{++}_L$. 
Comparison with Drell-Yan pair production 
processes shows that the present study  
extends the discovery reach of $\Delta^{++}$ at LHC considerably for most
of the allowed parameter space of the model.
The production of a single 
$\Delta^{++}_L$ in $pp$ collisions was
also investigated in ref. \cite{dicus} in the framework of Georgi-Machacek
model, which differs from the left-right symmetric model in some
phenomenologically important points, e.g. it has a new complex triplet
with Y=0 and it is assumed that there are no Majorana type couplings of
doubly charged Higgs to leptons.

Let us briefly summarize the present experimental limits on the mass
$M_{\Delta}$ and the possibly non-diagonal couplings $h_{ij}$ of the
doubly charged scalar particle  (see   \cite{hRll} and references
therein). The most stringent constraint comes from the upper limit for
the flavour changing decay $\mu\to\overline eee$:
\be h_{e\mu}h_{ee}<3.2\times 10^{-11}\;{\rm GeV}^{-2}\cdot
M_{\Delta^{++}} ^2.
\label{hemu}\ee 
From the Bhabha scattering  cross section at SLAC and
DESY the following bound  on the $h_{ee}$ coupling was established: 
\be h_{ee}^2 \lsim 9.7 \times 10^{-6}\; {\rm GeV }^{-2}\cdot
M_{\Delta^{++}} ^2  .
\label{hee}
\ee
 For $h_{ee}$=0.6, for example, the mass of the doubly charged
boson should obey
$M_{\Delta^{++}} \gsim 200$ GeV, but for smaller couplings much lighter
doubly charged scalars are still allowed. For the coupling $h_{\mu\mu}$
the extra contribution to $(g-2)_{\mu}$ yields the limit
\be h^2_{\mu\mu}\lsim 2.5\cdot 10^{-5}\; {\rm GeV}^{-2}\cdot 
M_{\Delta^{++}} ^2, 
\label{hmumu}\ee  and the muonium transformation to antimuonium converts
into a limit 
\be h_{ee} h_{\mu\mu } \lsim 5.8 \cdot 10^{-5}\; {\rm GeV}^{-2}\cdot
M_{\Delta^{++}}^2.
\label{heehmumu}\ee  From non-observation of the decay $\mu\to e\gamma$
follows
\be h_{e\mu }h_{\mu\mu}  \lsim 2 \cdot 10^{-10}\; {\rm GeV}^{-2}\cdot
M_{\Delta^{++}}^2.
\ee From the condition of vacuum stability one can derive upper bounds
on the couplings independent on the triplet Higgs mass
\cite{MohapatraStab}: $h_{ee},\; h_{\mu\mu }
\lsim 1.2$. 

The paper proceeds as follows. In the next section we will briefly
describe the left-right symmetric model.  Section 3 contains our results
for the production and decay of
$\Delta^{++}_{L,R}$, as well as a discussion on the background.
Conclusions are presented in Section 4.

\section{Description of the left-right symmetric model}

In this section we will present the basic structure of the \ssu
left-right symmetric model. Quarks and leptons are assigned to the
doublets of the gauge groups $SU(2)_{L}$ and $SU(2)_{R}$ according to
their chirality:
\bea
 &\Psi_L = {\matr{c} \nu_e \\ e^-\ematr_L} = (2,1,-1),\;\; &\Psi_R =
{\matr{c} \nu_e \\ e^-\ematr_R} = (1,2 ,-1),\nonumber\\ &Q_L = {\matr{c}
u \\ d\ematr_L}= (2,1,\third ),\;\; &Q_R = {\matr{c} u \\ d\ematr_R} =
(1,2 ,\third ),
\eea
 and similarly for the other families. The minimal set of fundamental
scalars, if the theory is symmetric under the $L\leftrightarrow R$
transformation, consists of the
 following Higgs multiplets:
\be
\begin{array}{c} {\dis\Phi =\matr{cc}\phi_1^0&\phi_1^+\\\phi_2^-&\phi_2^0
\ematr = (2,2,0)},
\\[15pt]
 {\dis\Delta_L =\matr{cc}\Delta_L^+&\sqrt{2}\Delta_L^{++}\\
\sqrt{2}\Delta_L^0&-\Delta_L^+
\ematr = (3,1,2)},\\[15pt]
{\dis\Delta_R=\matr{cc}\Delta_R^+&\sqrt{2}\Delta_R^{++}\\
\sqrt{2}\Delta_R^0&-\Delta_R^+\ematr = (1,3,2)}.
\end{array}
\ee They transform according to $\dis\Phi\to U_L\dis\Phi U_R^{\dagger}$,
$\Delta_L
\to U_L\Delta_L U_L^{\dagger}$ and $\Delta_R
\to U_R\Delta_R U_R^{\dagger}$, where $U_{L(R)}$ is an element of 
$SU(2)_{L(R)}$. The vacuum expectation value of the bidoublet $\dis\Phi$
is given by
\be
\begin{array}{c}
{\dis\langle\Phi\rangle=\frac1{\sqrt{2}}\matr{cc}\kappa_1&0\\0&\kappa_2\ematr.}
\end{array}\label{kappa}
\ee  This breaks the Standard Model symmetry \sulu. It generates masses
to fermions through the Yukawa couplings $\bar\Psi_L^i(f_{ij}\Phi +
g_{ij}\tilde\Phi)\Psi_R^j + h.c.$ and $\bar Q_L^i(f^q_{ij}\Phi +
g^q_{ij}\tilde\Phi)Q_R^j + h.c.$, where
$\tilde\Phi=\sigma_2\Phi^*\sigma_2$. 

The vacuum expectation values of the scalar triplets  are denoted by
\be
\begin{array}{c}  {\dis\langle\Delta_{L,R}\rangle
=\frac1{\sqrt{2}}\matr{cc}0&0\\v_{L,R}&0
\ematr.}
\end{array}\label{vlr}
\ee The right-triplet $\dis\Delta_R$ breaks the $SU(2)_R\times
U(1)_{B-L}$ symmetry to
$U(1)_Y$, and at the same time the discrete $L\leftrightarrow R$
symmetry, and it yields a Majorana mass to the right-handed neutrinos,
as was discussed  in  Introduction. It was also mentioned that the
vacuum expectation value
$v_L$ of the left-triplet is  quite tightly bound by the $\rho$
parameter. One has ($\kappa^2=\kappa_1^2+\kappa_2^2$)
\be
\rho=\frac{M_{W_L}^2}{\cos^2\theta_WM_{Z_1}^2}\simeq
\frac{1+2v_L^2/\kappa^2}{1+4v_L^2/\kappa^2},
\ee  and the experimental result \cite{PDG}  $\rho = 1.0004\pm 0.003$
then  implies
$v_L\lsim 9$ GeV, a small value compared with $ \kappa\simeq 250$ GeV. 

The Higgs potential describing the mutual interactions of $\Phi$,
$\Delta_R$ and
$\Delta_L$ is in general quite complicated containing a great number of
parameters. The most general potential for these fields  is given in
ref. \cite{Deshpande}. 
There are severe constraints on the model parameters, 
the most crucial ones for our study concern 
$v_L$ and flavour changing neutral currents (FCNC).
It was argued in  ref. \cite{Deshpande} that in phenomenologically 
consistent models either $v_L$ is exactly zero or a certain
combination of the potential parameters, $(2 \rho_1 - \rho_2),$ should
vanish identically. Since there is no fundamental principle
requiring  $v_L\equiv 0$ we assume the latter possibility
and study how to   probe the value of $v_L$ in 
single $\Delta^{++}_L$ production at LHC. In this case
one expects the splitting between $\Delta^{++}_L$ and $\Delta^{++}_R$
masses to be large leading to the relatively light $\Delta^{++}_L.$

Unlike in the SM, in the LR model there are FCNC interactions mediated
by some neutral Higgs fields which are certain superpositions of the neutrals
members of the bidoublet
$\Phi$.  
In order  to suppress FCNC one must require  the Higgs
potential to be such that in the minimum $\kappa_1\ll \kappa_2$ or
$\kappa_1\gg
\kappa_2$. This requirement has the consequence that the $W_L,W_R$ mixing
$\zeta$ is necessarily small, because $\zeta\simeq
(g_L/g_R)2|\kappa_1\kappa_2| /|v_R|^2$. We will assume 
$\kappa_2=0$, which leaves $W_L$ and $W_R$ as unmixed physical particles.

 The masses of the charged gauge bosons are  given in the case of no
mixing and
$v_L=0$ by the exact formulae
\bea {M_{W_L}^2}&=&\frac{1}{4}g_L^2\kappa_1^2,\\
{M_{W_R}^2}&=&\frac{1}{4}g_R^2(2v_R^2+\kappa_1^2).
\eea If the $L\leftrightarrow R$ symmetry is implemented, the gauge
couplings
$g_L$ and $g_R$ should be equal ($g_R=g_L\simeq 0.64$). If no such
symmetry is assumed, the internal consistency within  the model requires
nevertheless
$g_R\gsim 0.55\, g_L$ \cite{Boris}.  In order to satisfy the lower mass
limits of the new weak bosons $W_R$ and $Z_R$, the vev of the
right-handed triplet, $v_R$, should be considerably larger than 
$\kappa_1$ . Using the experimental value of the ordinary weak boson
$M_{W_L}=81$ GeV and the Tevatron lower bound $M_{W_R}\gsim 650$ GeV and
assuming
$g_R\simeq g_L = 0.64$, we find $v_R\gsim 5.6\; \kappa_1\simeq 1.4 $ TeV.

The  see-saw mass matrix of neutrinos is given by
\be
   M = \left(   \begin{array}{cc} m_L & m_D \\ m_D^T & m_R  \end{array} 
\right).
\label{seesaw}\ee
 The entries are $3\times 3$ matrices given by $m_D=(f\kappa_1 +
g\kappa_2)/\sqrt{2}$,
$m_L = h_L v_L$ and
$m_R=h_R v_R$. The mass of the charged lepton is given by
$m_l=(f\kappa_2 + g\kappa_1)/\sqrt{2}$, and therefore if $f$ and
$g$ are comparable, one has $m_D\simeq m_l$. Unless there is an
extraordinary hierarchy among the couplings, one has
$m_L\ll m_D\ll m_R$. In this case the approximate masses of the 
Majorana states that diagonalize the neutrino Lagrangian are given by 
  $ m_{\nu_1} \simeq m_D^Tm_R^{-1}m_D$ and $m_{\nu_2} \simeq  m_R$.

Ignoring the mixing between families and considering the matrix
(\ref{seesaw}) to present mixing of left-handed and right-handed
neutrino of one family, one obtains the mass eigenvalues
$m_{\nu_1}\simeq m_D^2/m_R$ and $m_{\nu_2}\simeq m_R$, and the mixing
angle $\eta$ between these states is given by
\be
   \tan 2\eta = \frac{2m_D}{m_R}.
\label{eta}\ee  The mixing of left- and right-handed neutrinos  is thus
in general very small, and one can identify $\nu_1$ with $\nu_L$ and
$\nu_2$ with $\nu_R$ as a good approximation.

The mass of the heavier neutrino $\nu_2$, the "right-handed" neutrino,
is related to the mass of $W_R$ via
\be m_{\nu_2}\simeq\frac{h_R}{g_R}M_{W_R}.
\ee  Most naturally the heavy neutrino and the heavy weak boson would
have roughly the same mass, but depending on the actual value of the
Yukawa coupling constant $h_R$ the neutrino may  also be lighter or
somewhat heavier than $W_R$.

\section{Signals of doubly charged Higgs production at LHC}

\subsection{ Production of doubly charged Higgses at LHC} 

\noindent We will assume that the doubly charged Higgs bosons  
$\Delta^{++}_{L,R}$ are  light enough
to be produced as  real particles at LHC, and we will study separately
their production and subsequent decay modes, concentrating mainly on the
right-triplet $\Delta_R^{++}$. However, in the cases when the situation 
in the left-handed sector differs from the right-handed one we will present
the differences. 
In $pp$-collisions $\Delta_{L,R}^{++}$ cannot
be produced in interactions of quarks because of the charge conservation.
However, it can be produced through $W_{L,R}W_{L,R}$ fusion, either by virtual
or real
$W_{L,R}$'s, depending on the mass of $\Delta_{L,R}^{++}$. The $WW$
fusion is not the only possibility, since also the physical  singly
charged Higgs scalars ($\kappa_2=0$ assumed)
\bea
 &&h^+ = \frac{1}{\sqrt{1+\frac{\kappa_1^2}{2v_R^2}}} (\Phi_1^+ +
\frac{\kappa_1}{\sqrt{2} v_R}
\Delta_R^+),\label{delta+}\\ 
&&\delta^+ = \frac{1}{\sqrt{1+\frac{2v_L^2}{\kappa_1^2}}}
(\Delta_L^+ +
\frac{\sqrt{2}v_L}{\kappa_1} \Phi_2^+ ), 
\label{phi+}\eea have  tree level couplings both to the doubly charged
Higgses and the quarks.

  Another  production mechanism of $\Delta^{++}_{L,R}$ at $pp$
 collider is via $\gamma, Z_L, Z_R$ exchange, i.e. 
$q\overline q\to \gamma^*,Z_R^* \to \Delta^{++}_R\Delta^{--}_R$ and 
$q\overline q\to \gamma^*,Z_L^* \to \Delta^{++}_L\Delta^{--}_L.$
The latter process has been studied at SSC energy in ref. \cite{grifols}. 
While for low values of $\Delta^{++}_{L,R}$
mass the cross section of the pair production is comparable with, or
may  even exceed, that of the 
$WW$ fusion reactions, its  kinematical reach is much lower.
 Indeed,  we are going to see that with
the present experimental constraints on model parameters 
the $\Delta^{++}_{L,R}$ 
mass reach at LHC in the Drell-Yan processes is roughly half of the 
one in $WW$ fusion processes.

\begin{figure}
\leavevmode
\begin{center}
\mbox{\epsfxsize=13.8truecm\epsfysize=3.2truecm\epsffile{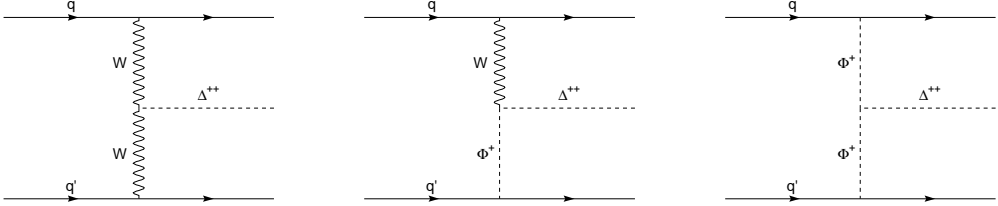}}
\end{center}
\caption{\label{graphs}Feynman graphs contributing to the production of
doubly charged Higgs at LHC.}
\end{figure}

The Feynman graphs for the  production processes of single 
$\Delta_R^{++}$ are depicted in Fig.\ref{graphs}, where $\phi^+$ stands
for the scalars $h^+$ and
$\delta^+$. Let us  investigate the relative importance of the 
amplitudes. The $WW$ fusion vertex $W_R^-W_R^-\Delta_R^{++}$ follows
from the gauge invariant kinetic term
$(D^{\mu}\Delta_R)^{\dagger}D_{\mu}\Delta_R$ as a result of the
spontaneous breaking of the left-right symmetry, and it is given by
\be
\frac{1}{\sqrt{2}} g_R^2 v_R W_R^-W_R^-\Delta_R^{++}.
\label{WWD}\ee 
The higher the symmetry breaking scale $v_R$ is, the
stronger the coupling is, but on the other hand the heavier $W_R$ is.
The corresponding vertex for the left-triplet
$\Delta^{++}_L$ is proportional to the small quantity $v_L$, which 
suppresses the fusion production of $\Delta^{++}_L.$  However, in this case
the lightness of $W_L$ boson enhances the production rate and, as
explained in Section 2, $\Delta^{++}_L$ is expected to be somewhat lighter
than $\Delta^{++}_R.$ If $v_L=0$ then the Drell-Yan process is the only 
possibility to produce the left-triplet Higgs boson.

\input{epsf.sty}

The $\phi^-\phi^-\Delta_R^{++}$ coupling arises from the following
quadratic terms of the scalar potential (we use the notation of
ref. \cite{Deshpande}):
\be
\beta_1\Tr (\Phi\Delta_R\Phi^{\dagger}\Delta_L^{\dagger}) + \beta_2\Tr
(\tilde\Phi\Delta_R\Phi^{\dagger}\Delta_L^{\dagger}) + \beta_3\Tr
(\Phi\Delta_R\tilde\Phi^{\dagger}\Delta_L^{\dagger}),
\ee
 which yields the couplings
\be
\frac{1}{\sqrt{2}} v_L\Delta_R^{++}(\beta_1\Phi_1^-\Phi_2^-
-\beta_2\Phi_1^-\Phi_1^- -\beta_3\Phi_2^-\Phi_2^-).
\label{PPD}\ee This contributes much less to the $\Delta_R^{++}$
production than the $WW$ fusion (\ref{WWD}) for several reasons. First,
the dimensionful  coupling $v_L$ is tiny compared with that of the
$W_R^-W_R^-\Delta_R^{++}$ vertex $v_R$. Second, the couplings of the
bidoublet fields $\Phi_1$ and $\Phi_2$ to quarks are Yukawa couplings
and  hence proportional to the  mass ratio $m_q/M_{W_L}$, which is small
compared with the gauge coupling $g_R$ appearing in the $WW$ fusion
amplitude. Furthermore, in realistic models the coupling constants
$\beta_i$ are necessarily small \cite{Deshpande}, 
otherwise one would face  a serious
fine tuning problem. In the case of $\delta^+$ there is still an extra
suppression due to the fact that the bidoublet component appears in
$\delta^+$ with just a small weight of
${\sqrt{2}v_L}/{\kappa_1}\lsim 0.04$ (see Eq. (\ref{phi+})).

The $\Delta_R^{++}W_R^-\phi^-$ vertex in the graphs of Fig.1.  derives
from the kinetic term $(D^{\mu}\Delta_R)^{\dagger}D_{\mu}\Delta_R$,
which yields  the coupling
\be ig_R W_{R\mu}^-(\Delta_R^-\partial^{\mu}\Delta_R^{++}-
\Delta_R^{++}\partial^{\mu}\Delta_R^{-}).
\ee This amplitude is possible for $h^-$, but not for $\delta^-$ which
does not have a right-triplet component. For $h^-$ the process is
suppressed, in addition to the small Higgs coupling with 
quarks, also due to the small weight
${\kappa_1}/{\sqrt{2} v_R}$ of the $\Delta_R^-$ component.

We can conclude that the $W_RW_R$ fusion process dominates the
production of single
$\Delta^{++}_R$ in $pp$ collisions, and the Higgs
exchange diagrams can be safely ignored (except in the unprobable case
that $W_R$ is orders of magnitude heavier than the singly charged
Higgses). The same is also true for the left-triplet Higgs boson
production process.

We have performed  calculations of the $WW$ fusion process in Fig.1. 
without making any simplifying assumptions
taking into account the three particle final state phase space.
To obtain numerical values for the cross sections and final state 
distributions we have convoluted the functions over the initial state quarks
momentum spectra using the default MRS-G set of the parton distributions
of CERN Library program PDFLIB \cite{pdflib}. All numerical integrations 
have been performed by the integration routine VEGAS \cite{vegas} which
ensures high accuracy of the results.

\begin{figure}
\leavevmode
\parbox[t]{15cm} {
\begin{center}
\mbox{\epsfxsize=9truecm\epsfysize=9.truecm\epsffile{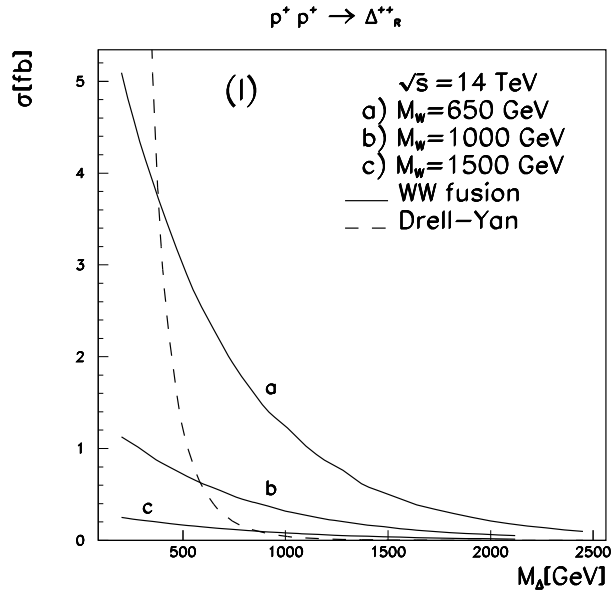}}
\end{center}}
\parbox[t]{15cm}
{\begin{center}
\mbox{\epsfxsize=9truecm\epsfysize=9.truecm\epsffile{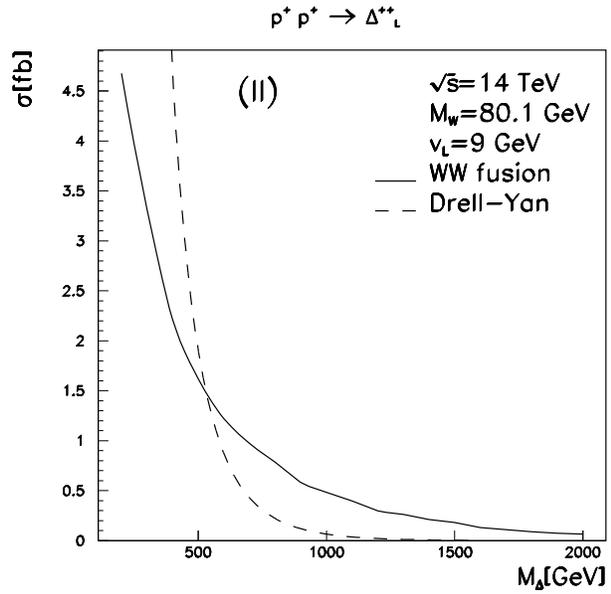}}
\end{center}}
\caption{\label{prodcs}Production cross sections for $\Delta^{++}_{R,L}$
as  functions of doubly charged Higgs mass. }
\end{figure}

The production cross sections for $\Delta^{++}_{L,R}$ in $WW$ fusion
are presented in  Fig.\ref{prodcs}. by bold lines.   
In Fig.\ref{prodcs}. (I) the cross section of 
$\Delta^{++}_R$ production is plotted 
for three different values of $W_R$ mass as a function
of $\Delta^{++}_R$ mass. For low $\Delta^{++}_R$ masses the cross section 
is a quite rapidly falling function of $M_{W_R},$ 
e.g. going from 650 GeV to 1.5 TeV, the cross section for 
$M^{++}_{\Delta_R}=200$
GeV decreases  more than an order of magnitude. If light $\Delta^{++}_R$
will be detected then this sensitivity can be used to obtain indirect 
information about $W_R$ mass. For heavy $\Delta^{++}_R,$ however, 
the cross section depends on $M_{W_R}$ rather weakly which allows  one
to probe large Higgs masses. With the present lower value $M_{W_R}=650$ GeV,
 the designed LHC luminosity of 100 fb$^{-1}$ per year and assuming
that ten events are needed for discovery, $\Delta^{++}_R$
as heavy as 2.4 TeV can be found at LHC. 
We have also calculated the Drell-Yan pair production cross 
section of $\Delta^{++}_R$  (assuming  $Z_R$ to be very heavy)
and presented it by dashed line in Fig.\ref{prodcs}. (I).   
For $M_{W_R}\lsim 1$ TeV and 
$M_{\Delta_R}^{++}\gsim 500$ GeV the $WW$ fusion cross section exceeds
the Drell-Yan one which falls below the discovery limit if 
$M_{\Delta_R}^{++}\gsim 1$ TeV.

If $v_L $ vanishes, the
production of $\Delta_L^{++}$ in $W_L^+W_L^+$ fusion is not possible due
to the proportionality of the corresponding coupling to $v_L$
analogously to Eq. (\ref{WWD}). The single production by other mechanisms is
too small to be observed as explained previously. If $v_L $ differs from
zero, also $\Delta_L^{++}$ production is possible. For the allowed
 vev $v_L=9$ GeV the $WW$ fusion production cross section is shown in
Fig.\ref{prodcs}. (II) by the bold line.  In this
case the free parameters to be tested are the vev $v_L$ and the mass of
$\Delta_L^{++}$ (the cross section scales as $v_L^2$). 
With the chosen value of $v_L$ the
production cross section is comparable with the 
$\Delta^{++}_R$ production one but it falls faster with the Higgs mass. 
The discovery limit of $\Delta^{++}_L$  is as high as 1.75 TeV.
In this case the Drell-Yan cross section, also presented in 
Fig.\ref{prodcs}. (II), 
is a few percent bigger than for the right-triplet Higgs due to 
additional $Z_L$ contribution but, as seen in figure, its discovery potential
is the same.  

\begin{figure}
\leavevmode
\parbox[t]{15cm} {
\begin{center}
\mbox{\epsfxsize=10truecm\epsfysize=10.truecm\epsffile{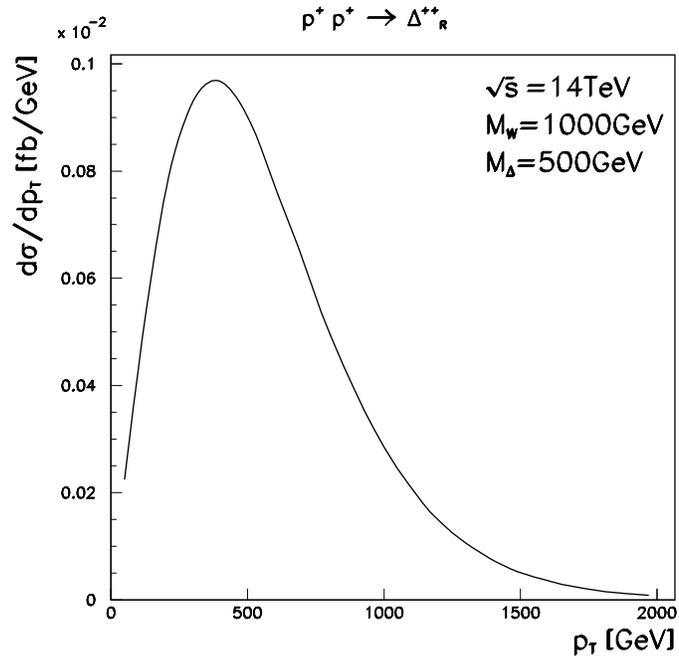}}
\end{center}}
\parbox[t]{15cm}{
\begin{center}
\mbox{\epsfxsize=10truecm\epsfysize=10.truecm\epsffile{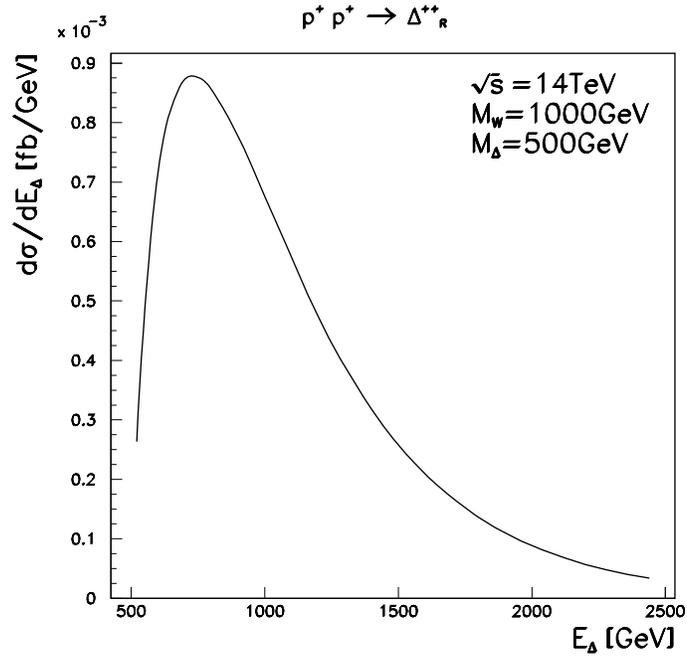}}
\end{center}}
\caption{\label{dcsang}Distributions of $p_T$ and $E_{\Delta_R}$ for
doubly charged Higgs production.}
\end{figure}

Especially for background considerations it will be interesting to study
the  angular distribution of $\Delta_R^{++}$. In Fig.\ref{dcsang}. the
$p_T$ and $E_{\Delta_R}$ distributions are shown for
$m_{\Delta_R^{++}}=$500 GeV and $M_{W_R}=$1 TeV. Comparing the
distributions it is seen that the doubly charged Higgses are  mostly
transverse. As the angular distribution of the decay products of
$\Delta^{++}_R$ is flat, the loss of events in small forward angles is 
not essential.

\subsection{Decay of the triplet Higgs}

In the lowest order the doubly charged scalar $\Delta_R^{++}$  can decay
via the following channels:
\bea
\Delta_R^{++} &\to & l^+l'^+, \label{Dll}\\
 &\to & W_R^+W_R^+, \label{DWW}\\
 &\to & W_R^+h^+, \label{DWh}\\
 &\to & W_R^+W_R^+\delta^0, \label{DWWd}\\
 &\to & h^+h^+,	 \label{Dhh}\\
 &\to & h^+h^+h^0, \label{Dhhh}\\
 &\to & h^+h^+\delta^0,	\label{Dhhd}\\
 &\to & W_R^+ h^+ Z_1,	 \label{DWhZ1}\\
 &\to & W_R^+ h^+ Z_2	 \label{DWhZ2}.
\eea 
Here the neutral scalars $h^0$ and $\delta^0$  are mixtures of the
real components of the states $\Phi_1^0$ and $\Delta_R^0$ so that in a
good approximation $h^0\simeq
\Phi_1^{0r}$ and
$\delta^0\simeq \Delta_R^{0r}$.  The charged Higgses $h^+$ and
$\delta^+$ were defined in Eqs. (\ref{delta+}) and (\ref{phi+}). 

Let us consider the relative magnitudes of the different decay channels.
If   there is a reasonable degeneracy between $\Delta_R^{++}$ and the
other members of the right-triplet, i.e. the singly charged scalar $h^+$
having a right-triplet component and the predominantly right-triplet
neutral scalar $\delta^0$, the modes (\ref{DWWd}) to (\ref{DWhZ2}) are
likely to be kinematically suppressed or disallowed.  The channel
$\Delta_R^{++} \to  W_R^+ h^+ Z_1$, while  not necessarily kinematically
suppressed, is in general quite small due to the smallness of $Z_L,Z_R$
mixing. This would leave us with the channels
$\Delta_R^{++}
\to l^+l'^+$, $\Delta_R^{++} \to W_R^+W_R^+$,
$\Delta_R^{++} \to  W_R^+h^+$. 

Kinematically most favoured is the decay to a same-sign lepton pair, 
$\Delta_R^{++} \to l^+l'^+$, for which the decay width is given by
\be
\Gamma(\Delta_R^{++} \to l^+l'^+) = \frac{h_{R,ll'}^2}{8\pi}
M_{\Delta_R^{++}}
\simeq 4\; {\rm GeV}\cdot h_{R,ll'}^2 (\frac{M_{\Delta_R^{++}}}{100\;
{\rm GeV}}),
\ee where the lepton masses are neglected.  In Introduction  the present
bounds on the coupling constants $h_{R,ll'}$ were given, and for the
mass of
$\Delta_R$ in the range we will assume it to lie,
$M_{\Delta_R^{++}}\gsim 0.5$ TeV, the diagonal couplings $h_{R,ee}$,
$h_{R,\mu\mu}$ and $h_{R,\tau\tau}$ can according to these bounds be as
large as {\cal{O}}(1). In the case  $l=l'=e$ or $l=l'=\mu$ we will have
an unambiguous signature of a same-sign lepton pair with no missing
energy. In the case   $l=l'=\tau$  the final state would include either
two same-sign leptons  ($e^+e^+$, $\mu^+\mu^+$ or $e^+\mu^+$) with
missing energy or transverse pions plus missing energy. The even more
spectacular signature of a same-sign electron-muon pair with no missing
energy would be suppressed due to smallness of the non-diagonal coupling
$h_{R,e\mu}$ as the bound (\ref{hemu}) seems to indicate (or otherwise
$h_{R,ee} \ll h_{R,e\mu}$).

The decays  $\Delta_R^{++} \to W_R^+W_R^+$ and
$\Delta_R^{++} \to  W_R^+h^+$  may also be kinematically disfavoured as the
mass of the heavy boson $W_R$ is in general of the same order of
magnitude as the mass of the Higgs triplet. If this is the case the
decay to a like-sign lepton pair is the dominant mode in the lowest
order. Let us, however, assume for the time being that $W_R$ is light
enough for these processes to proceed and compare the magnitude of their
widths with $
\Gamma(\Delta_R^{++} \to l^+l'^+)$.

The rate of $\Delta_R^{++} \to W_R^+W_R^+$ is given by 
\bea
\Gamma(\Delta_R^{++} \to W_R^+W_R^+) = \frac{g_R^2
M_{\Delta_R^{++}}}{8\pi} &\sqrt{1 -
4(\frac{M_{W_R}}{M_{\Delta^{++}}})^2}(1-(\frac{M_{W_L}}{M_{W_R}})^2) (3
\frac{M_{W_R}^2}{M_{\Delta^{++}}^2}\nonumber\\
&+\frac{M_{\Delta^{++}}^2}{4M_{W_R}^2}-1).
\eea
The $ \Delta^{++}_RW_R^-W_R^-$ vertex is proportional to the vev  $v_R$,
and thus this decay channel need not be suppressed like the
corresponding channel of the left-triplet, where the coupling is
proportional to the small or possibly vanishing vev $v_L$ of
$\Delta_L^0$. Actually, for $M_{W_R}=0.5$ TeV, the branching ratio of 
$\Delta_R^{++} \to W_R^+W_R^+$ exceeds that of $\Delta_R^{++} \to l^+l^+$
when 
$M_{\Delta^{++}}\gsim 2$ TeV, assuming that the Majorana coupling
$h_{R,ll}$ is of the order of one. 
If the energy of $\Delta^{++}_R$ is not high enough for decay into a pair of
real $W_R^+$'s, a decay channel including one or two virtual $W_R^+$'s
might still be important.  
We have studied this possibility, but the lepton channel is then
the dominating one, unless it is for some reason totally suppressed.
If $h_{R,ll'}\sim 0$, the branching ratio to virtual $W_R^+$'s is
close to the branching ratio of $\Delta_R^{++}\rightarrow W_R^+h^+$.

The detection signal of the decay $\Delta_R^{++} \to W_R^+W_R^+$ would
be a hard like-sign lepton pair with missing energy, assuming that the
two neutrinos produced along with the charged leptons are the light
Majorana neutrinos $\nu_1$ or heavy Majorana neutrinos $\nu_2$ with a
mass $m_2\ll M_{W_R}$.  It can be distinguished from the similar final
state of the $\Delta_R^{++}\to
\tau^+\tau^+$ channel on grounds of its different angular distribution.
If the heavy neutrino produced in the $W_R$ decay has a large mass, it
will decay in the detector via
$\nu_2\to lW^*_R\to l + two\; jets$. The decay products are likely to be well
separated, but nevertheless in the LHC environment this would
constitute  a much less background free signal than purely leptonic
final states. The same is true for the case where 
$W_R$'s are reconstructed in  hadronic channels due to $W_R\to two\;
jets$. 

Note that for non-vanishing $v_L$ the decay channel 
$\Delta_L^{++} \to W_L^+W_L^+$ can be the dominant one for the 
left-triplet Higgs $\Delta_L.$ In this case one has to tag   four jets
coming from $W_L$ decays and reconstruct the invariant 
mass of $\Delta_L^{++}.$

The decay width of $\Delta_R^{++} \to  W_R^+h^+$ is given by 
\be
\Gamma(\Delta_R^{++} \to  W_R^+h^+)=\frac{g_R^2
M_{\Delta_R^{++}}}{16\pi}\frac{M_{W_L}^2
  M_{\Delta_R^{++}}^2}{M_{W_R}^4}
\lambda^{3/2}\left( 1,
\left( \frac{M_{W_R}}{M_{\Delta_R^{++}}}\right)^2,
\left(\frac{M_h}{M_{W_R}}\right)^2\right),
\ee 
where $\lambda$ is the usual phase space suppression function. For
small values of the $\Delta_R^{++}$ mass, this width is considerably
smaller than the width of the lepton pair channel, assuming that $h^+$
has a mass of the same order of magnitude as the mass of $W_R$ or
larger. The $h^+$ scalar decays  via its $\Delta_R^+$ component to
$l^+_R\nu_R$ and via its $\Phi^+$ component mainly to a top
quark-antiquark pair.

To gain an idea of the relative magnitudes of the widths of the three
main decay modes discussed above, consider some representative cases.
For $  M_{\Delta_R^{++}}= 1$ TeV, $M_{W_R}=0.5$ TeV, $M_h=0.2$ TeV and
$m_{\nu_2}=1$ TeV   one has
$\Gamma(\Delta_R^{++} \to l^+l'^+)= 13$ GeV, 
$\Gamma(\Delta_R^{++} \to
W_R^+W_R^+)= 0$ GeV and
 $\Gamma(\Delta_R^{++} \to  W_R^+h^+)= 0.3 $ GeV.   
For $  M_{\Delta_R^{++}}=
2$ TeV, $M_{W_R}=0.5$ TeV, $M_h=0.5$ TeV and $m_{\nu_2}=1$ TeV  the
widths are 80 GeV, 90
 GeV and
 4 GeV, respectively. 

In the following  we will consider more closely the detection of
$\Delta^{++}_R$ via its leptonic decay and discuss the background from
SM sources.

\subsection{ Backgrounds from the Standard Model and the Left-Right
Model} 

\noindent We shall argue that a like-sign lepton pair as an
experimental signal of
$\Delta_R^{++}$ production has a very low background from other
processes. If the pair of like-sign electrons or muons is seen, as is
the case with  relatively light or very heavy $\Delta_{L,R}^{++}$'s
(Fig.3.), the amount of missing transverse energy is tiny. Since all
possible background channels from the SM contain missing energy due to
neutrinos \cite{BCHP} in order to preserve lepton number,  there is no 
background from the SM in the events of interest.

In the case of the left-right model, the background of the same order in 
perturbation theory is the "neutrinoless double beta decay" process, in
which two same sign $W_R$'s fuse to a like-sign lepton pair via
exchanging a Majorana neutrino in t-channel.  Using EPA distribution for
longitudinal and transverse distributions of $W_R$'s \cite{JL} inside
colliding protons, we calculated the upper limit for this background.
Although the contribution from individual process is larger if the 
$W_R$ is longitudinal \cite{HHMR}, the beams contain mostly  transverse
gauge bosons  \cite{JL}. It turns out that the contribution from the
transverse $W_R$'s is the dominant one. For the masses we have used, the
neutrinoless double beta decay contribution is
${\cal {O}}(10^{-4})$ fb. This is negligible compared with the signal
cross section.

The mass of the doubly charged Higgs can be reconstructed from the same
charge leptons in the signal cases. As discussed in the previous
section, the peak in the case of leptonic  decay modes is relatively
narrow (for $M_{\Delta_R^{++}} = 1$ TeV and $M_{W_R}=$ 0.5 TeV one has
typically $\Gamma(\Delta_R^{++} \to l^+l'^+)= {\cal O}(10)$ GeV). Thus
it should be clearly visible in the signal cases.
 
\begin{figure}[t]
\leavevmode
\begin{center}
\mbox{\epsfxsize=10truecm\epsfysize=10truecm\epsffile{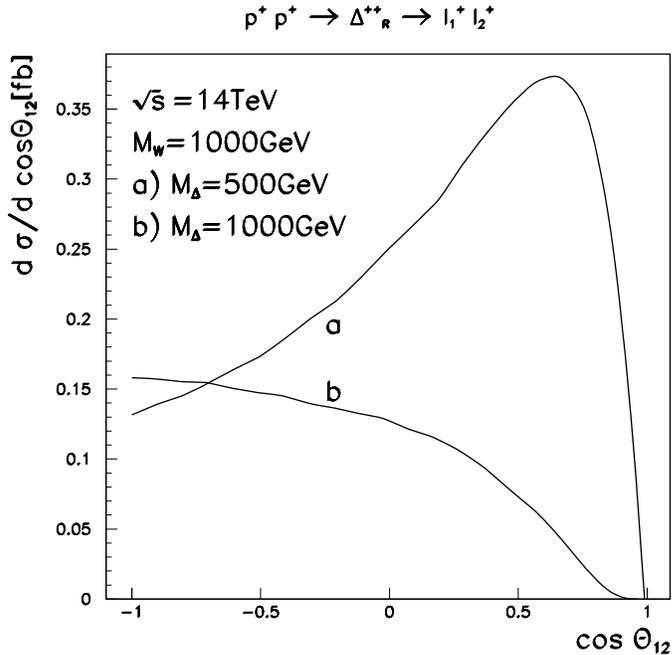}}
\end{center}
\caption{\label{theta12} Opening angle distribution between leptons in
$pp\to\Delta^{++}_R\to l^+ l^+$ in the $pp$ center of mass frame.  }
\end{figure}

Let us finally note that the opening angle between the two leptons
strongly depends on the mass of $\Delta_R^{++}$, as is demonstrated in
Fig.\ref{theta12}. In the case of a small mass
$\Delta_R^{++}$'s are produced in average with a larger three-momentum
than in the case of a large mass, making the decay products more boosted.
\input{epsf.sty}

\section{Conclusions and discussion}

We have studied the single production of doubly charged scalars
$\Delta_{L,R}^{++}$ through $W_{L,R}^+W_{L,R}^+$ fusion in $pp$ 
collisions at LHC in the framework of the left-right symmetric
electroweak model. 
Since the relevant couplings involved are rather general,
most of our results are also applicable in other models 
with scalar bileptons.
The production cross section of the right-triplet Higgs
$\Delta_R^{++}$  depends on its mass as well as on the mass of $W_R.$ 
An interesting feature of the cross section is that the $W_R$ mass dependence
is relatively weak for large $M_{\Delta_{R}^{++}}$
which allows  one to probe large Higgs masses. 
For representative mass ranges 200 GeV $\lsim
M_{\Delta_R^{++}}\lsim$ 1000 GeV and 650 GeV $\lsim
M_{W_R^{++}}\lsim$ 1000 GeV the cross section is 0.3 to 5 fb. With
the anticipated LHC luminosity of $10^{34} \;{\rm cm}^{-2}{\rm s}^{-1}$ this
would mean of the order of a few hundred events per year. 
If the mass of $W_R$ is
close to its present lower limit 650 GeV and we require 10 events for 
discovery then one year of effective LHC run 
will cover $M_{\Delta_R^{++}}$ up to 2.4 TeV.  
The Drell-Yan pair production cross section exceeds the $WW$ fusion one 
only for $M_{\Delta_{R}^{++}}\lsim 0.5$ TeV and falls below the observable 
limit if $M_{\Delta_{R}^{++}}\gsim 1$ TeV.

Single $\Delta_{L}^{++}$ production can take place if $v_L$ is non-zero.
With the maximal presently allowed value $v_L=9$ GeV the LHC discovery
reach during one year run extends up to 1.75 TeV, 
while the corresponding limit in Drell-Yan process is 1 TeV. 
Note that in this
case $\Delta_{L}^{++}$ is expected to be somewhat lighter than 
$\Delta_{R}^{++}$ which makes the $WW$ fusion process especially 
interesting to search for.   
In conclusion, with the present constraints on the model parameters
the  $\Delta^{++}$  masses achievable in $WW$ fusion processes exceed by 
a factor of two the ones testable in Drell-Yan pair production.

In the accessible mass range the decay of
$\Delta_R$ is dominated by the channel
$\Delta_R^{++}\to l^+l^+$, providing an excellent signal. This decay
provides interesting information, because its amplitude measures the
strength of the lepton number violating Yukawa coupling, which plays a
central part in the see-saw mechanism of neutrino masses.  Depending on
the respective masses, the decay modes 
$\Delta_R^{++}\to W_R^+W_R^+$ and 
$\Delta_R^{++}\to W_R^+h^+$ may also contribute. 
In the case of the left-triplet Higgs boson the decay
$\Delta_L^{++}\to W_L^+W_L^+,$ which can be detected via $W_L$ hadronic
decay channels, 
may dominate over the decay
$\Delta_L^{++}\to l^+l^+.$ 

\bigbreak

\noindent {\bf Acknowledgements} 
We thank R. Vuopionper\"a for discussions.
J.M. gratefully acknowledges a grant
from Jenny ja Antti Wihurin rahasto and M.R. a post doctoral grant from
Spanish Ministry of Science and Education.  This work has been supported
by the Academy of Finland, Turun Yliopistos\"a\"ati\"o, and CICYT, grant
AEN-93-0234.


\begin{thebibliography}{99}

\bibitem{cuy} For an overview see, 
F. Cuypers and S. Davidson, hep-ph/9609487, 
submitted to Phys. Rep.

\bibitem{pati} J.C. Pati and A. Salam, Phys. Rev. D 10 (1974) 275;\\
R.N. Mohapatra and J. C. Pati, Phys. Rev. D 11 (1975) 566, 2558;\\ G.
Senjanovic and R. N. Mohapatra, Phys. Rev. D 12 (1975) 1502;\\ R. N.
Mohapatra and R. E. Marshak, Phys. Lett. 91 B (1980) 222.
 
\bibitem{seesaw} M. Gell-Mann, P. Ramond and R. Slansky, in {\it
Supergravity}, eds. P. van Niewenhuizen and D. Z. Freedman (North
Holland 1979);\\ T. Yanagida, in Proceedings of {\it Workshop on Unified
Theory and Baryon Number in the Universe}, eds. O. Sawada and A.
Sugamoto (KEK 1979).

\bibitem{sun} See e.g., B.T. Cleveland et al., in the {\it Proceedings
of the 16th International Conference on Neutrino Physics and
Astrophysics}, eds. A. Dar, G. Eilam and M. Gronau, Eilat, Israel 29 May
- 3 June 1994, Nucl. Phys. B (Suppl.) 38 (1995) 47; Y. Suzuki, ibid. 54;
J.N. Abdurashitov et al., ibid. 60; P. Anselmann et al., ibid. 68.

  
\bibitem{atmos}K.S. Hirata et al., Phys. Lett. B 280 (1992) 146;\\ D.
Casper et al., Phys. Rev. Lett. 66 (1993) 2561.
 
 
\bibitem{cobe}G.F. Smoot {\it et al.}, Ast. J. 396 (1992) L1.


\bibitem{CDF} F. Abe {\it et al.}, CDF Collaboration, Phys. Rev. Lett. 68
(1992) 1463.

\bibitem{D0} S. Abachi {\it et al.}, D0 Collaboration, Phys. Rev. Lett.
76 (1996) 3271.

\bibitem{PDG} Particle Data Group, {\it Review of Particle Physics},
\PR{D 54} (1996) 1.

\bibitem{rizzoap95} 
P. Langacker and S. Uma Sankar, Phys. Rev. D 40 (1989) 1569; \\
T. Rizzo, Phys. ReV. D 50 (1994) 325, and {\it
ibid.} 5602; \\
G. Barenboim, J. Bernabeu, J. Prades and M. Raidal, hep-ph/9611347,
submitted to Phys. Rev. D.


\bibitem{NLC} T. Rizzo, Phys. Lett. 116 B (1982) 23; D. London, G.
Belanger and J.N. Ng, Phys. Lett. B 188 (1987) 155;
 J. Maalampi, A. Pietil\"a and J. Vuori,  Nucl. Phys. B 381 (1992) 544,
and Phys. Lett. B 297, 327 (1992);
 J. Maalampi and  A. Pietil\"a, Z. Physik C 59 (1993) 257; C. A. Heusch
and P. Minkowski, Nucl. Phys. B 416, 3 (1994);
 K. Huitu, J. Maalampi and M. Raidal, Nucl. Phys. B 420 (1994) 449, and
Phys. Lett. B 328 (1994) 60;
 J.Gluza and M. Zra\l ek, Phys. Rev. D 52 (1995) 6238; J.Gluza and M.
Zra\l ek, Phys. Lett. B 362 (1995) 148;
 P. Helde, K. Huitu, J. Maalampi and M. Raidal, Nucl. Phys. B 437 (1995)
305; A. Pietil\"a and J. Maalampi, Phys. Rev. D 52 (1995) 1386.

\bibitem{gunionDelta} J.F. Gunion, Int. J. Mod. Phys. A 11 (1996) 1551. 

\bibitem{acco} E. Accomando and S. Petrarca, Phys. Lett. B 323 (1994)
212.

\bibitem{grifols} J.A. Grifols, A. Mendez and G.A. Schuler,
Mod. Phys. Lett. A 4 (1989) 1485; \\
J.F. Gunion, J. Grifols, A. Mendez, B. Kayser, and F.
Olness, Phys. Rev. D 40 (1989) 1546.

\bibitem{dicus} D. Dicus and R. Vega, Nucl. Phys. B 329 (1990) 533.


\bibitem{hRll} M.L. Swartz, Phys. Rev. D 40 (1989) 1521; \\
 M. Lusignoli and S. Petrarca, Phys. Lett. B 226 (1989) 397; \\ 
R. Mohapatra, Phys. Rev. D 46 (1992) 2990.

\bibitem{MohapatraStab} R. Mohapatra, Phys. Rev. D 34 (1986) 909.

\bibitem{Deshpande} N.G. Deshpande, J.F. Gunion, B. Kayser, and F.
Olness, Phys. Rev. D 44 (1991)  837.

\bibitem{pdflib} H. Plothow-Besch,  Comp. Phys. Comm. 75 (1993) 396; \\
H. Plothow-Besch,  Int. J. Mod. Phys. A 10 (1995) 2901.

\bibitem{vegas} G.P. Lepage, J. Comp. Phys. 27 (1978) 192.


\bibitem{Boris} M. Cvetic, P. Langacker and  B. Kayser,
 Phys. Rev. Lett. 68 (1992) 2871.


\bibitem{BCHP} V. Barger, K. Cheung, T. Han, R.J.N. Phillips,
\PR{D}42 (1990) 3052. 

\bibitem{JL} J. Lindfors, Z. Phys. C 28 (1985) 427.

\bibitem{HHMR} See P. Helde et al. in reference \cite{NLC}.


 
\end{thebibliography}
\end{document}